\documentclass[conference]{IEEEtran}
\IEEEoverridecommandlockouts
\usepackage{cite}
\usepackage{amsmath,amssymb,amsfonts}
\usepackage{algorithmic}
\usepackage{graphicx}
\usepackage{textcomp}
\usepackage{xcolor}
\usepackage{hyperref}

\def\BibTeX{{\rm B\kern-.05em{\sc i\kern-.025em b}\kern-.08em
    T\kern-.1667em\lower.7ex\hbox{E}\kern-.125emX}}
\begin{document}

\title{Link Prediction of Artificial Intelligence Concepts using Low Computational Power
}

\author{\IEEEauthorblockN{Francisco Valente}
\IEEEauthorblockA{\textit{no affiliation}\\
Leiria, Portugal \\
paulo.francisco.valente@gmail.com}
}

\maketitle

\begin{abstract}

This paper presents an approach proposed for the “Science4cast” 2021 competition, organized by the Institute of Advanced Research in Artificial Intelligence, whose main goal was to predict the likelihood of future associations between machine learning concepts in a semantic network. The developed methodology corresponds to a solution for a scenario of availability of low computational power only, exploiting the extraction of low order topological features and its incorporation in an optimized classifier to estimate the degree of future connections between the nodes. The reasons that motivated the developed methodologies will be discussed, as well as some results, limitations and suggestions of improvements.
\end{abstract}


\section{Introduction}

This manuscript is related to the individual solution developed for a competition organized by the Institute of Advanced Research in Artificial Intelligence  (IARAI): “Science4cast” \footnote{More information about the competiton can be found at \url{https://www.iarai.ac.at/science4cast/}}, an official competition within the 2021 IEEE BigData Cup Challenges. The purpose of the competition was the prediction of the association of artificial intelligence (AI) or machine learning (ML) concepts, i.e., to forecast if two given AI/ML concepts will appear together in the title or abstract of a future scientific paper. For that purpose, the organizers provided a semantic network of 64719 vertices (AI/ML concepts) that characterizes the AI scientific literature from 1994 to 2017, where a link represents the association of two ML concepts. The final goal was to predict the 2020 likelihood of links between vertices that were unconnected at the end of 2017. 
\par
In order to solve this task, the author had access only to very low computational power tools\footnote{For the sake of interest, the computational tool mainly used for this task were two laptop computers with the following characteristics: (1) Intel Core i7-8550U processor with 8GB RAM, and (2) Intel Core i7-4710HQ processor with 12GB RAM.}. This had limited and guided all the steps of the followed approach, as it will be further described and detailed. This report presents then a solution developed for such a constrained scenario, explaining how it was overcome and also how the methodology could be improved by having access to better computational tools.

\section{Possible approaches and followed solution}

This competition is related to the “link prediction problem”, where the goal is to predict if two unconnected vertices/nodes of a graph at a time t0 will be connected (have an edge/link) at a time t1. More specifically, to estimate the likelihood of such connection. In order to obtain such forecasting, the methodologies typically rely on the development of a classification model based on some types of features: 

\begin{itemize}
    \item Node-attribute features: those correspond to node-specific properties. For the scenario of ML concepts, it could include, for example, the number of papers where such concepts are mentioned, the authors of those articles, the count of citations.
    \item Topological features: those are generated based on the local network structure, i.e., they take into account the nodes' neighboring structure.
    \item Latent features: those are obtained through embedding methods that derive lower-dimensional vectors from the graph. Some examples include matrix factorization, PageRank, SimRank, Katz, DeepWalk, node2vec, LINE. 
\end{itemize}

For this competition, node features were not available, and so only topological and latent features could be considered. As mentioned in the Introduction section, this work was largely constrained by the required computational power and the limited tools the author had available. 
\par
Latent feature methods are typically high-order algorithms, i.e., they consider the entire semantic network in order to generate the embedding vectors. As abovementioned, the network is composed of 64719 vertices, which makes it a very large network. Moreover, this is a time-dependent graph network, where new links are continuously created. In other words, if we divide the semantic network by years, there is a different graph for each one of the years from 1994 to 2017. Considering the limited computational tools, even for a single year, the network is so large that it was infeasible to compute those embedding methods.
\par
Therefore, only topological features were extracted in the proposed solution, as those features are low-order ones: they only take into account the close neighbors of the pair of nodes.
\par
For the prediction of edges in 2020, there was a total of 24 different graphs (1994-2017). In a simple link-prediction scenario, the features extracted from a single graph could be inputted into a classification model to predict the link-likelihood of unconnected nodes. In this case, if we consider features for more than one year (and thus more than one graph), we must combine all the values of a given feature extracted for the selected years. For that purpose, some algorithms may be created to directly model the time variance of the values of the features. For example, a variation of a recurrent neural network was thought of. However, the simplest approach (both in terms of methodology and computational power) seems to be to consider each year as a new feature. More specifically, if we consider data from 5 years, and we have a set of 10 types of features, a total of 50 features (1 feature per year) will be combined into a single set and used to create the classification model. In this way, the author do not directly establish the time variance of each feature over the years (which is expected to be preferable), but he relies on the classifier to try to deal with such information.
\par
In short, the proposed solution considers a group of topological features extracted from a set of graphs related to each year, which are combined to train a classifier. The following section will explore each one of the steps considered to do so.

\section{Data from the semantic network}

As already mentioned, the semantic network has data from 1994 to 2017. In principle, the more years we take into account, the better the model we can obtain. However, again, for that purpose, high-performance computing is required. Therefore, it was necessary to limit the number of years to a short period. More specifically, the author considered a period of 3 years (which was also the baseline of the competition's tutorial) as a good compromise.
\par
Once only 3 years are considered, it was necessary to select whose years from the 24 available ones would be selected to train the classification model. In principle, the most reasonable choice would be to select the last 3 years, because if there is a concept drift, i.e., a variation in the relationship between input and output data over the years, then it is expected that the most recent years predict better the next years. For this scenario, as the goal is to predict the 2020 links and 2017 is the last labelled year, that would mean to select as training data the 2012-2014 graphs with the 2017 label.
\par
However, analyzing the results of the tutorial of the competition, such a trend in the concept drift was not verified. In that tutorial, a set of 15 (topological) features were extracted from the years 2009-2011 of the semantic network to train a (neural network) prediction model, considering the 2014 label. That model was then applied to predict the edges of the semantic network at the end of different years. More specifically, the following evaluations were assessed in the tutorial:

- Train prediction: 2014 link prediction of the vertices unconnected in 2011, using data from the 2009-2011 semantic network. In this case, data were divided into two groups: 90\% for training (in this manuscript it would be called "2014-train prediction"), and 10\% hold-out group for testing (in this manuscript it would be called "2014-test prediction"),
\par
- Validation prediction ("2017 prediction"): 2017 link prediction of the vertices unconnected in 2014, using data from the 2012-2014 semantic network.
\par
- Evaluation prediction ("2020 prediction"): 2020 link prediction of the vertices unconnected in 2017, using data from the 2015-2017 semantic network. This corresponds to the final goal of the competition.

The area under the ROC curve (AUC) was computed for all those predictions. As a baseline, the "2014-train prediction" had an AUC of $\approx$0.7640. The AUC values for the unseen data were:

- "2014-test prediction": AUC$\approx$0.7853
\par
- "2017 prediction": AUC$\approx$0.7227
\par
- "2020 prediction": AUC$\approx$0.8798
 
\par

Considering a period of 3 years only, it seems there is no evident tendency for the concept drift. The "2014-test prediction" is similar to the "2014-train" one, which is expected if the prediction model is not overfitted. For the 2017 predictions, there is a significant drop in the predictive performance, which suggests the model generated using the 2009-2011 data is not so suitable to estimate the 2017 links. However, the same model applied to analyze the 2020 data performs even much better than the training one. 
\par
Therefore, even only three time-points are considered in this analysis, it suggests a high and non-linear variance in the concept drift as, for a short period of 6 years, the predictive performance drops $\approx$6\% (2014$\xrightarrow{}$2017), and then it highly improves $\approx$15\% (2017$\xrightarrow{}$2020). 
\par
In fact, as this is a time-dependent scenario, it would be expected that if there is a drop in the predictive performance it means the trained model lost its predictive ability for more recent years, and so the 2020 result is very incoherent. Further analysis of both this variance over the years and the respective data would be required to understand why such drifts occur. This may be of interest in further studies to get more insights related to the growth of this semantic network, and thus improve the estimation of future edges as well.
\par
In short, for the sake of this competition, considering only the three analyzed time-points, it is suggested that the initial training period used in the tutorial (2009-2011 data with the 2014 label) is a very good one to predict the 2020 links. In other words, to train a classification model, the topological features were extracted for the years 2009, 2010 and 2011.
\par
As there are 64719 nodes, there is a total of 2094242121 possible edges. From those, only 2278611 were connected (around 0.001\%). In other words, there are more than 2000 million edges that could be considered for training analysis. Surely, this is unfeasible without high-performing computing. Therefore, it was necessary to select a much sparser subset of edges for analysis. It was observed that the large majority of those vertices had a zero degree (had no links with other nodes) in the 2009-2011 period, and thus if we select randomly a given number of nodes, the topological features would have a value of zero for the majority of samples. In order to overcome it, the logical step was to select only edges with nodes with a non-zero degree. For that purpose, the author followed the baseline selection of the competition's tutorial of using a minimal degree of 10, as it allowed to significantly reduce the set of possible edges and having topologically "rich" nodes (many neighbors). In fact, there were 94043755 edges whose both vertices had a minimal degree of 10 in 2011, from which $\approx$99.6\% were unconnected. Thus, it still requires significant computational power. In order to speed up all the steps (namely features' extraction, and classifiers' optimization and training), an undersampling was executed. From that set of pairs, 523721 pairs were linked 3 years later (at the end of 2014). In order to have good representations of both positive (nodes connected at 2014) and negatives (nodes not connected at 2014), while keeping a significantly representative part of the total dataset, all the positive pairs were selected, and the same number of negative samples was randomly selected. Therefore, the training dataset selected to be used in this work had a total of 1047442 samples. 
\par
While this was the approach followed for the competition, the author acknowledges "post-competition" that this methodology could probably be improved. For the evaluation dataset, some nodes may have a degree lower than 10. Therefore, a (probably) better approach could be to select two groups of nodes: one with degree$\geq$10 as performed, and another with 1$\geq$degree$<$10, and then join their data into a single dataset.

\section{Extracted topological features}

For the extraction of low order topological features, the author analyzed the literature to look for features used in related problems. For that purpose, works from several domains were considered, such as association of concepts in quantum physics \cite{Krenn2020}, link-prediction in scientific papers citation \cite{Liu2019}, prediction of protein interactions\cite{Kovacs2019}, similarity between genes \cite{Bass2013}, among others \cite{Liben2003, Srilatha2016}. Based on that, 12 types of features that exploit the neighborhood properties of the nodes to provide were considered:

\begin{enumerate}
    \item Degree centrality (DC): number of neighbors of each node in the pair, i.e., how many links a given vertice has. [2 features per pair, one for each node]
    \item Total number of neighbors (TN): sum of the individual degrees of the nodes in the pair. [1 feature per pair]
    \item Common neighbors index (CN): number of vertices that are neighbors of both nodes of the pair. [1 feature per pair]
    \item Jaccard similarity coefficient (JC): ratio of common neighbors between the two nodes of the pair in relation to all the neighbors of both nodes. [1 feature per pair]
    \item Simpson similarity coefficient (SC): ratio of common neighbors between the two nodes of the pair in relation to the lowest degree of the pair. [1 feature per pair]
    \item Geometric similarity coefficient (GC): ratio of the square of the number of common neighbors of the two nodes in the pair in relation to the product of the individual degrees. [1 feature per pair]
    \item Cosine similarity coefficient (CC): square root of the geometric coefficient. [1 feature per pair]
    \item Adamic-Adar index (AA): sum of the inverse logarithmic of the degree of the common neighbors of the nodes in the pair. [1 feature per pair]
    \item Resource allocation index (RA): sum of the inverse degree of the common neighbors of the nodes in the pair. [1 feature per pair]
    \item Preferential attachment index (PA): product of the degree centrality of each node in the pair. [1 feature per pair]
    \item Average neighbor degree (AD): average degree of the neighbors of each node in the pair. [2 features per pair, one for each node]
    \item Clustering coefficient (CI): fraction of the number of triangles through each node in the pair in relation to its degree; it measures the degree to which the neighbors of the node tend to form a cluster (complete graph).  [2 features per pair, one for each node]
\end{enumerate}

All the described features are then low order ones, i.e., for each pair of nodes, only the knowledge of its neighbors (first-order) and its neighbors’ neighbors (second-order) is required. Some of those features (namely the indices like JC, SC, GC, CC, AA, RA, PA) are sometimes used to directly estimate the likelihoods of links between nodes, i.e., to inform about the future connectivity of unconnected nodes without a machine learning procedure. However, here, they are incorporated simply as topological features.
\par
The semantic network of this competition is an undirected graph, i.e., there is no source and target nodes for each link. The features that are obtained for each node of the pair individually (DC, AD, CI) may contribute to the prediction of the future connection between those nodes, and then they are included. However, a more sophisticated approach that allows to get that information without indirectly assuming source and target nodes would probably be better. 
\par
As mentioned in the previous section, those features were obtained for a 3 years period. More specifically, for the 2020 predictions, data is extracted for the 2015-2017 period, and the training dataset's features are obtained for the 2009-2011 period. In other words, each feature is extracted three times, one for each year. Therefore, for each pair of concepts, a total of 45 features are obtained:

\begin{itemize}
    \item DC[v1,y1], DC[v2,y1], DC[v1,y2], DC[v2,y2], DC[v1,y3], DC[v1,y3]
    \item TN[y1], TN[y2], TN[y3]
    \item CN[y1], CN[y2], CN[y3]
    \item JC[y1], JC[y2], JC[y3]
    \item SC[y1], SC[y2], SC[y3]
    \item GC[y1], GC[y2], GC[y3]
    \item CC[y1], CC[y2], CC[y3]
    \item AA[y1], AA[y2], AA[y3]
    \item RA[y1], RA[y2], RA[y3]
    \item PA[y1], PA[y2], PA[y3]
    \item AD[v1,y1], AD[v2,y1], AD[v1,y2], AD[v2,y2], AD[v1,y3], AD[v1,y3]
    \item CI[v1,y1], CI[v2,y1], CI[v1,y2], CI[v2,y2], CI[v1,y3], CI[v1,y3]
\end{itemize}
\color{black}

Where v1 is one of the vertices of the pair, v2 is the other vertex of the same pair, y1 is the last year of the data period (2011 for training, 2017 for evaluation), y2 is the middle year of the data period (2010 for training, 2016 for evaluation), y3 is the oldest year of the data period (2009 for training, 2015 for evaluation).
\par
Univariate statistical analysis tests (mutual information and chi-squared) were performed to study the individual relationship between each one of the features and the true label, and all of them were found to be highly significant and so they were all considered for the next steps. 
\par
Finally, the final training dataset was standardized, and the corresponding standardization was applied to the final evalation dataset.  

\section{Principal component analysis of correlated features}

As can be noticed through the explanation of each of the topological features considered in the previous section, some of them are expected to be highly correlated as they are derived similarly. The Pearson correlation between all features was assessed, and some groups where the features are all very correlation were found:

\begin{itemize}
    \item Group 1: features DC[v1,y1], DC[v1,y2], DC[v1,y3] (correlations in the interval 95-99\%).
    \item Group 2: features DC[v2,y1], DC[v2,y2], DC[v2,y3] (correlations in the interval 97-99\%).
    \item Group 3: features TN[y1], TN[y2], TN[y3] (correlations in the interval 96-99\%).
    \item Group 4: features CN[y1], CN[y2], CN[y3] (correlations in the interval 96-99\%).
    \item Group 5: features JC[y1], JC[y2], JC[y3], GC[y1], GC[y2], GC[y3], CC[y1], CC[y2], CC[y3] (correlations in the interval 63-95\%, but most correlations $\geq$70\%).
    \item Group 6: features SC[y1], SC[y2], SC[y3] (correlations in the interval 77-91\%).
    \item Group 7: features AA[y1], AA[y2], AA[y3], RA[y1], RA[y2], RA[y3], PA[y1], PA[y2], PA[y3]
 (correlations in the interval 74-97\%, but most correlations $\geq$80\%).
\end{itemize}

Therefore, based on that, the author proposed to perform a dimensionality reduction through Principal Component Analysis (PCA). The goal is to get a smaller set of variables that are linear combinations of the original features. This allows to group very correlated features while reducing the total number of features, which may contribute to a classification model that generalizes better for unknown data.
\par
The PCA technique was then applied to each one of the above groups, for the training dataset (2009-2011 data). For each of those seven groups, the explained variance ratio of the first principal component (eigenvector) was the following one:

\begin{table}[h!]
\centering
\caption{Explained variance ratio of PCA's first principal component, for each group in the training dataset.}
\label{tab:pca_variance}
\begin{tabular}{|l|c|}
\hline
        & \textbf{Explained variance (\%)} \\ \hline
Group 1 & 98.75\%                 \\ \hline
Group 2 & 99.24\%                 \\ \hline
Group 3 & 96.68\%                 \\ \hline
Group 4 & 99.07\%                 \\ \hline
Group 5 & 93.94\%                 \\ \hline
Group 6 & 92.06\%                 \\ \hline
Group 7 & 98.43\%                 \\ \hline
\end{tabular}
\end{table}

As observable in Table \ref{tab:pca_variance}, the first principal component of the PCA of each one of the groups explains more than 92\% of their features' variance, and for four of those groups it even explains more than 98\%.
\par
The first principal components alone already explain a very large part of the groups' variance. Therefore, each of those groups (both in training and evaluation datasets) was replaced by their projection on the PCA's first principal component of the corresponding group (PCA-Gx), as obtained with the training dataset. Thus, the final group of 19 variables to be used to train a classification model was the following one:

\begin{itemize}
    \item PCA-G1, PCA-G2, PCA-G3, PCA-G4, PCA-G5, PCA-G6, PCA-G7, AD[v1,y1], AD[v2,y1], AD[v1,y2], AD[v2,y2], AD[v1,y3], AD[v1,y3], CI[v1,y1], CI[v2,y1], CI[v1,y2], CI[v2,y2], CI[v1,y3], CI[v1,y3]
\end{itemize}

\section{Training of classification model}

Having the final set of features, the last phase of the proposed solution was the training of a classification model to predict the link likelihood of previously unconnected vertices. For that purpose, four types of classifiers were considered: elastic-net logistic regression (LR), k-nearest neighbors (KNN), random forest (RF), and multilayer perceptron (MLP). Support vector machine (SVM) was also considered, but it was found to require a lot of time and computational power, and thus it was discarded due to the mentioned limitations.
\par
In a first step, the hyperparameters of those classifiers were optimized using a random search, through a 5-fold cross-validation, using the training dataset. For that purpose, to speed up the computation, only a part of the 1047442 samples of the dataset was considered. More specifically, 100000 samples were used, selecting randomly 50000 positive and 50000 negative samples. The selected parameters were the ones that maximized the mean AUC of the 5 folds.
\par
The following parameters were considered for the classifiers' optimization:

\begin{table}[h!]
\centering
\caption{Parameters selected to be optimized using a random search. The names of the parameters are the ones used in the package \textit{sklearn}, which was considered to train the classifiers.}
\label{tab:clf_parameters}
\begin{tabular}{|c|c|}
\hline
\textbf{Classifier} & \textbf{Optimized parameters} \\ \hline
LR         & l1\_ratio, alpha          \\ \hline
KNN        & n\_neighbors, weights     \\ \hline
RF  & \begin{tabular}[c]{@{}c@{}}n\_estimators, max \_depth, min\_samples\_split, \\ min\_samples\_leaf\end{tabular}        \\ \hline
MLP & \begin{tabular}[c]{@{}c@{}}hidden\_layers\_sizes, activation, alpha, learning\_rate, \\ learning\_rate\_init, max\_iter\end{tabular} \\ \hline
\end{tabular}
\end{table}

\par
For RF and MLP classifiers, as they have a larger number of parameters to be optimized, a second round of random search optimization was performed, around the parameters found to be the best ones in the first round.
\par
From the four classifiers, random forest was found to be the one with the best hold-out predictive performance (AUC$\approx$0.87 - mean of the five cross-validation folds).

An ensemble strategy, based on the combination of the predictions of the four classifiers, weighted by their training AUC, was also considered. For that task, as the estimated probabilities of the KNN and RF are not inherently meaningful (and then they may be over and/or underestimated), their prediction was first calibrated (using a logistic regression model), and only then the results of the four classifiers were combined. However, that approach was not found to significantly improve the AUC performance of the RF. Therefore, considering an “Occam's razor” reasoning, the optimized random forest was selected as the classifier to be used to predict the link likelihoods of the 2020 semantic network.
\par
Having the final classifier, a new round of training was performed. As described, for the hyperparameters optimization a sparser dataset (100000 samples) was used. In this new round, considering the selected parameters, the random forest was trained using all the 1047442 samples initially retrieved. That final trained model was then applied to the evaluation dataset of the competition.

\section{Result for the 2020 link prediction}

The proposed solution (optimized Random Forest trained on the 1047442-samples dataset) achieved a predictive performance of AUC$\approx$0.8925 in the evaluation dataset. As a comparison, the RF trained with the same parameters but on the sparser set of 100000 samples got AUC$\approx$0.8882. This suggests that an increase in the ability to predict new links may be obtained considering more data, emphasizing the importance of access to more computational power.

\section{Best model' parameters and code}

In this section, the author presents some additionally details of the final approach. More specifically the random forest optimized parameters and the PCA first component formulas:

\begin{table}[h!]
\centering
\caption{Selected parameters for the final Random Forest prediction model.}
\label{tab:rf_best_param}
\begin{tabular}{|c|c|}
\hline
\textbf{Parameter}          & \textbf{Optimized value} \\ \hline
n\_estimators       & 1250        \\ \hline
max\_depth          & None (default)                \\ \hline
min\_samples\_split & 7                \\ \hline
min\_samples\_leaf  & 5                \\ \hline
\end{tabular}
\end{table}

\begin{table}[h!]
\centering
\caption{Obtained first principal component using PCA for each one of the groups of correlated features. Note: for simplicity, in this table, only two decimal places of the eigenvectors are presented.}
\label{tab:pca_component}
\begin{tabular}{|c|c|}
\hline
 & \textbf{PCA's first principal component} \\ \hline
Group 1                                                                 & 
0.78*DC[v1,y1]+0.52*DC[v1,y2]+0.33*DC[v1,y3] \\ \hline
Group 2                                                                 & 
0.79*DC[v2,y1]+0.51*DC[v2,y2]+0.33*DC[v2,y3] \\ \hline
Group 3                                                                 & 0.90*TN[y1]+0.41*TN[y2]+0.17*TN[y3]      \\ \hline
Group 4                                                                 & 0.79*CN[y1]+0.52*CN[y2]+0.33*CN[y3]     \\ \hline
Group 5 &
  \begin{tabular}[c]{@{}c@{}}0.02*JC[y1]+0.02*JC[y2]+0.01*JC[y3]+\\ +0.82*GC[y1]+0.51*GC[y2]+0.25*GC[y3]+\\ +0.06*CC[y1]+0.04*CC[y1]+0.03*CC[y2]\end{tabular} \\ \hline
Group 6                                                                 & 0.74*SC[y1]+0.56*SC[y2]+0.36*SC[y3]     \\ \hline
Group 7 &
  \begin{tabular}[c]{@{}c@{}}2.04e-5*AA[y1]+1.14e-5*AA[y2]+6.05e-6*AA[y3]+ \\ +4.80e-7*RA[y1]+3.06e-7*RA[y2]+1.87e-7*RA[y3]+\\ +8.92e-1*PA[y1]+4.13e-1*PA[y2]+1.84e-1*PA[y3]\end{tabular} \\ \hline
\end{tabular}
\end{table}

Table \ref{tab:pca_component} may contribute to further analysis of which features contribute more to the first principal component of each group.

All the proposed solution was executed using Python scripts and publicly available packages, such as \textit{NumPy}, \textit{scikit-learn}, and \textit{NetworkX}. The code of the developed methodology is available at: \url{https://github.com/pfranciscovalente/science4cast_topological}.

\section{Conclusion}

As disclosed throughout 
the manuscript, the proposed approach was highly constrained by the available computational tools. In fact, the results suggest that without that limitation the results could be significantly improved, by considering more data. Furthermore, the use of more complex methods, namely embedding methodologies for the creation of latent features, and the use of node-attribute features, that were not available for this competition, are expected to significantly contribute to an improvement of the predictive performance. The methodology presented in this study is then a single procedure of a 3-part strategy that could be possibly followed to get the most from the semantic network, in other circumstances. Finally, while the developed approach does not lead to the most impressive predictive results compared to other participants of the competition, the analysis and discussion presented in this report may be useful for some considerations in future studies.

\section*{ACKNOWLEDGMENT}
\small
The author thanks Fábio Lopes for providing some additional computational support.

\bibliographystyle{IEEEtran}

\bibliography{science4cast}

\end{document}